\documentclass[aps, prl, reprint, twocolumn, superscriptaddress, amsmath, amssymb]{revtex4-2}
\usepackage{graphicx}% Include figure files
\usepackage{dcolumn}% Align table columns on decimal point
\usepackage{bm}% bold math
\usepackage{enumerate}
\usepackage{fancyhdr}
\usepackage{color}
\usepackage[normalem]{ulem}
\usepackage[squaren]{SIunits}
\usepackage[mathlines]{lineno}% Enable numbering of text and display math
\usepackage[colorlinks=true,linkcolor=magenta]{hyperref}
\hypersetup{
    citecolor=blue,
}

\begin{document}

\title{Measurements of the growth and saturation of electron Weibel instability in optical-field ionized plasmas}

\author{Chaojie Zhang}
\affiliation{Department of Electrical Engineering, University of California Los Angeles, Los Angeles, California 90095, USA}
\author{Jianfei Hua}
\affiliation{Department of Engineering Physics, Tsinghua University, Beijing 100084, China}
\author{Yipeng Wu}
\affiliation{Department of Electrical Engineering, University of California Los Angeles, Los Angeles, California 90095, USA}
\author{Yu Fang}
\affiliation{Department of Engineering Physics, Tsinghua University, Beijing 100084, China}
\author{Yue Ma}
\affiliation{Department of Engineering Physics, Tsinghua University, Beijing 100084, China}
\author{Tianliang Zhang}
\affiliation{Department of Engineering Physics, Tsinghua University, Beijing 100084, China}
\author{Shuang Liu}
\affiliation{Department of Engineering Physics, Tsinghua University, Beijing 100084, China}
\author{Bo Peng}
\affiliation{Department of Engineering Physics, Tsinghua University, Beijing 100084, China}
\author{Yunxiao He}
\affiliation{Department of Engineering Physics, Tsinghua University, Beijing 100084, China}
\author{Chen-Kang Huang}
\affiliation{Department of Electrical Engineering, University of California Los Angeles, Los Angeles, California 90095, USA}
\author{Ken A. Marsh}
\affiliation{Department of Electrical Engineering, University of California Los Angeles, Los Angeles, California 90095, USA}
\author{Warren B. Mori}
\affiliation{Department of Electrical Engineering, University of California Los Angeles, Los Angeles, California 90095, USA}
\affiliation{Department of Physics and Astronomy, University of California Los Angeles, Los Angeles, CA 90095, USA}
\author{Wei Lu}
\affiliation{Department of Engineering Physics, Tsinghua University, Beijing 100084, China}
\author{Chan Joshi}
\affiliation{Department of Electrical Engineering, University of California Los Angeles, Los Angeles, California 90095, USA}

\date{\today}

\begin{abstract}
The temporal evolution of the magnetic field associated with electron thermal Weibel instability in optical-field ionized plasmas is measured using ultrashort (1.8 ps), relativistic (45 MeV) electron bunches from a linear accelerator. The self-generated magnetic fields are found to self-organize into a quasi-static structure consistent with a helicoid topology within a few ps and such a structure lasts for tens of ps in underdense plasmas. The measured growth rate agrees well with that predicted by the kinetic theory of plasmas taking into account collisions. Magnetic trapping is identified as the dominant saturation mechanism.
\end{abstract}

% \pacs{V52.38.Kd, 52.65.-y, 41.75.Ht}% PACS, the Physics and Astronomy
                             % Classification Scheme.

\maketitle
Generation and amplification of magnetic fields in plasmas is a long-standing topic that is of great interest to both fundamental and applied physics. One well-known mechanism is the Weibel instability \cite{weibel_spontaneously_1959} that arises in plasmas with anisotropic electron velocity distribution (EVD). Magnetic fields are self-generated and rapidly amplified due to the self-organization of the microscopic plasma currents in anisotropic plasmas \cite{fried_mechanism_1959}. As the instability grows, the strength, wavevector spectrum and topology of the magnetic field evolve as a result of the continuous merging of currents \cite{fonseca_three-dimensional_2003}. The Weibel magnetic field may be responsible for seeding subsequent turbulence generation \cite{mondal_direct_2012} and dynamo amplification in galactic plasmas \cite{kulsrud_spectrum_1992}. Scenarios where the Weibel instability is thought to play a role include astrophysical phenomena such as gamma-ray bursts (GRBs) \cite{medvedev_generation_1999,lyubarsky_are_2006}, relativistic jets in active galactic nuclei (AGN) \cite{nishikawa_particle_2003}, neutrino winds \cite{silva_exact_2006} and collisionless shocks \cite{caprioli_cosmic-ray-induced_2013-1,blandford_particle_1987}; laboratory plasmas involved in inertial confinement fusion \cite{macchi_fundamental_2003,silva_role_2002,ren_global_2004}, laser-driven shocks \cite{fiuza_weibel-instability-mediated_2012,fox_filamentation_2013,huntington_observation_2015,swadling_measurement_2020,fiuza_electron_2020}, plasma-based particle acceleration \cite{su_stability_1987,yan_self-induced_2009,qiao_stable_2009}; matters at extreme conditions such as electron-positron \cite{yang_evolution_1994,fonseca_three-dimensional_2002} and quark-gluon plasmas \cite{arnold_apparent_2005}.

A particular type of Weibel instability driven by interpenetrating streams of beams or plasmas (also referred to as current filamentation instability, CFI) has been investigated in experiments by either passing a relativistic electron beam through a plasma \cite{allen_experimental_2012,raj_probing_2020}, driving locally heated  electrons through solid-density plasmas \cite{mondal_direct_2012,chatterjee_magnetic_2017,zhou_self-organized_2018,ruyer_growth_2020}, or by creating two counter propagating plasmas \cite{fox_filamentation_2013,huntington_observation_2015,swadling_measurement_2020,fiuza_electron_2020}. In these experiments, the characteristic filamentary magnetic field structures are purported to have been observed in either the electron beam itself \cite{allen_experimental_2012}, in the external probe beam of protons \cite{fox_filamentation_2013,huntington_observation_2015}, or by using optical polarimetry \cite{mondal_direct_2012,chatterjee_magnetic_2017,zhou_self-organized_2018}. However, few experiments have been able to capture the temporal evolution of the Weibel-CFI instability including its exponential growth, saturation and damping. On the other hand, in spite of being one of the earliest kinetic plasma instabilities had been discovered, the original concept of electron Weibel instability driven by a temperature anisotropy in a stationary, unmagnetized plasma (often referred to as thermal Weibel instability \cite{romanov_self-organization_2004}) has thus far not been observed to our knowledge.

In this Letter, we use an optical-field ionized (OFI) plasma to initialize a known anisotropic EVD and then make picosecond-time-resolved measurements of the growth, saturation and damping of the electron thermal Weibel instability.

\begin{figure}
\includegraphics[width=0.45\textwidth]{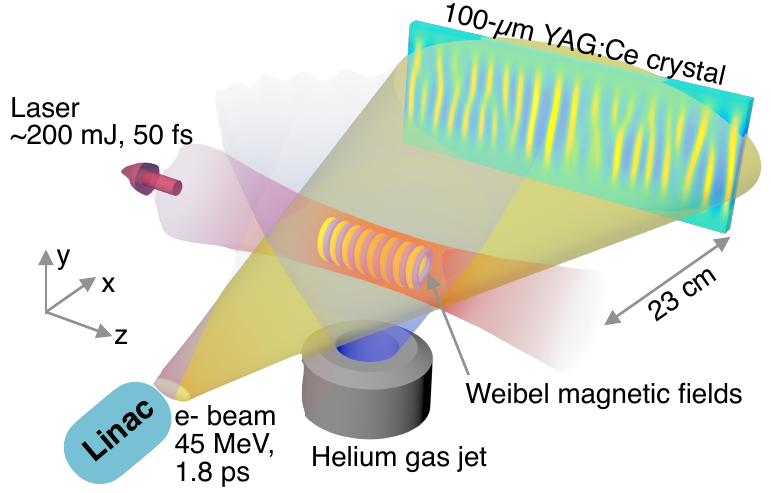} % this command will be ignored
\caption{The schematic drawing of the experimental setup.}
\label{fig1}
\end{figure}

The sketch of the experimental setup is shown in Fig. \ref{fig1}. A circularly polarized, ultrashort ($\tau\approx 50~\femto\second$, full width at half maximum, FWHM) Ti:Sapphire laser pulse ($\lambda_0\approx800~\nano\meter$) was focused to a $22\pm1~\micro\meter$ diameter spot to ionize helium gas emanating from a supersonic nozzle. The laser was intense enough ($\sim 2.5\times10^{17}~\watt/\centi\meter^2$) to rapidly ionize both electrons of the helium atoms during the risetime of the pulse through tunnel ionization \cite{ammosov_tunnel_1987} to generate a plasma with electron density in the range of $(0.3-1.5)\times10^{19}~\centi\meter^{-3}$ but without driving significant amplitude plasma wakes. The plasma was inferred to have a $\sim 200~\micro\meter$ diameter and the central $\sim 100~\micro\meter$ region was fully ionized (Supplemental Material). A relativistic, ultrashort electron bunch probe ($E\approx45$ MeV, $\Delta E/E\sim0.5\%$, $\epsilon_n\approx1~\rm mm\cdot mrad$, $\tau_{\rm FWHM}\approx1.8$ ps), containing $\sim30$ pC charge (too small to excite measurable wakefield) was orthogonally incident on the plasma (see Supplemental Material and \cite{du_generation_2013} for the beam configuration). Deflections caused by fast-oscillating fields (e.g., wakes) were not detected since $\tau_{\rm probe}\gg c/\omega_p$ \cite{zhang_capturing_2016,zhang_femtosecond_2017}. In other words, the probe electrons were deflected by the $\bf v\times B$ force exerted by the quasi-static magnetic fields in the plasma. These deflections translated into a density modulation that was captured by a thin scintillator screen and the subsequent imaging system (Supplemental Material).

The OFI plasma was intrinsically highly anisotropic with $A\equiv T_\perp/T_\parallel-1\gg 1$ due to the fact that electrons pick up energy from the laser field predominantly along the polarization direction when the laser is gone. Here $T_\perp$ and $T_\parallel$ are the effective transverse and longitudinal temperature, respectively, and $A$ is the plasma anisotropy. In addition to the large anisotropy, the initial transverse EVD of the OFI helium plasma consists of two concentric rings in the momentum space, as shown and measured in a recent experiment \cite{huang_initializing_2020}. In such a plasma there follows a hierarchy of kinetic instabilities that begins with largely electrostatic two-stream and the oblique current filamentation instabilities which have been measured with 100 fs resolution using Thomson scattering \cite{zhang_ultrafast_2019}. These instabilities not only reduce the plasma anisotropy rapidly from initially $>100$ to $\sim10$ in just one ps but also lead to approximately bi-Maxwellian plasma electrons with $T_\perp\approx 500~\electronvolt$ and $T_\parallel\approx 40~\electronvolt$ as observed in previous PIC simulations \cite{zhang_probing_2020}. Such a bi-Maxwellian EVD will be ideal for the growth of the Weibel instability which is predominantly an electromagnetic instability.

\begin{figure}
\includegraphics[width=0.45\textwidth]{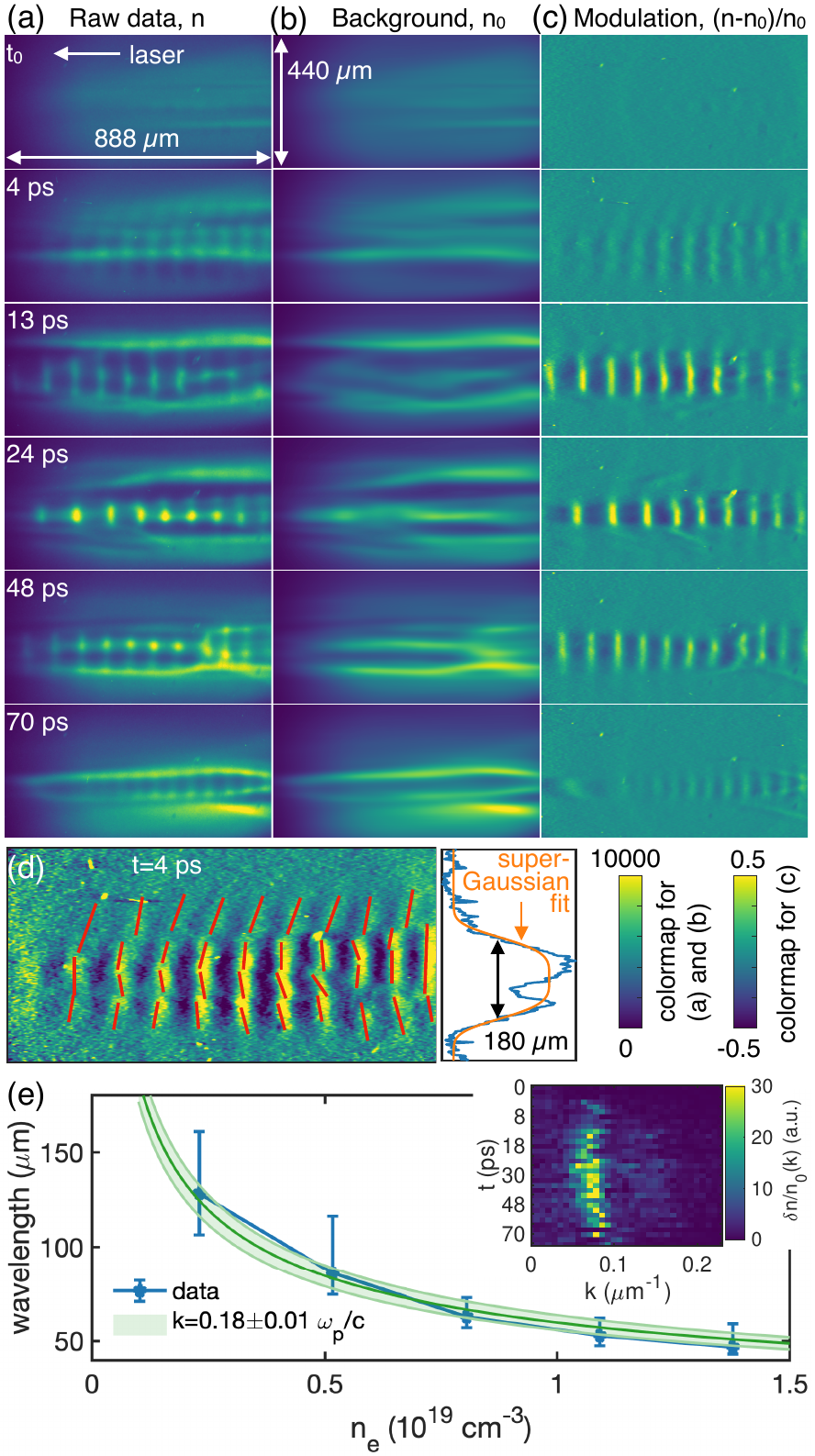} % this command will be ignored
\caption{(a) Time-resolved snapshots of the magnetic fields in OFI plasmas. The raw data of the beam profile recorded at increasing time delays is shown. (b) Reconstructed background (see text). (c) Relative density modulation of the electron beam. (d) The orientation of small magnetic field structures probed by the electron beam at 4 ps are marked by the red dashed lines. The magnetic field width integrated along $z$ is shown on the right. (e) Density dependence of the measured wavelength of the magnetic field. The error bar is inferred using the FWHM width of the integrated $k$ spectrum. The inset shows the time-resolved $k$ spectrum of $\delta n/n_0$.}
\label{fig2}
\end{figure}

Representative snapshots of the probe electron bunch after it had traversed the plasma taken at different delays with respect to the ionization laser are shown in Fig. \ref{fig2}(a). See Supplemental Material for the full dataset. Here $t_0$ is defined as the time when the electron beam overlaps with the laser at the interaction point, which was at the center of the camera view. The uncertainty of $t_0$ was estimated to be within 3 ps and the temporal resolution was $\sim2$ ps (Supplemental Material). The timing jitter between the ionizing laser pulse and the electron bunch was $\sim0.1$ ps \cite{lin_development_2018}.

The raw data in Fig. \ref{fig2}(a) show both large-scale and small-scale structures in the electron density images. To separate them, we average each snapshot line by line to smear out the vertically aligned small-scale structures and therefore what left is the contribution from large-scale fields in the plasma, as shown in Fig. \ref{fig2}(b). Possible sources of the large-scale structures and their effects on data analysis are discussed in Supplemental Material. The deflection of probe electrons by the Weibel magnetic fields ($B_y$ component, see below) sits on top of these large-scale structures. Therefore, we can treat each shot in Fig. \ref{fig2}(b) as the reconstructed single-shot background. In Fig. \ref{fig2}(c) we show the calculated relative density modulation $\delta n/n_0=(n-n_0)/n_0$, where $n$ is the raw density profile in Fig. \ref{fig2}(a) and $n_0$ is the reconstructed background in Fig. \ref{fig2}(b).

It can be clearly seen from Fig. \ref{fig2}(c) that the density modulation magnitude of the probe beam grows from some detection threshold level ($\delta n/n_0\gtrsim 0.03$ at $t_0$) to peak at about $t=24$ ps and then slowly drops in tens of ps. The predominant wavelength remains almost constant up to $t=48$ ps. Kinetic theory predicts that for a given temperature anisotropy $A$, Weibel instability starts growing with a wide range of unstable wavenumbers, $0<k<\sqrt{A}~\omega_p/c$, with the wavevector pointing along the cold temperature direction ($\hat z$) \cite{bret_multidimensional_2010}. As the instability grows, the initial broad $k$ spectrum narrows due to the self-organization (from merging of currents) of the magnetic fields. The expected initial broad $k$ spectrum was not captured in this experiment, nevertheless some fine modal structures are indeed visible at the early stage of the instability, as shown in Fig. \ref{fig2}(d). Previous work has predicted that for a plasma with similar bi-Maxwellian EVD, the magnetic fields eventually evolve into a quasi-static helicoid structure \cite{romanov_self-organization_2004,terekhin_helical_1999}, in the form of ${\bf B}\approx {\hat x}B_0 \cos kz + {\hat y}B_0 \sin kz$. Probe electrons deflected by ${\bf B}_y=\hat{y}B_0\sin kz$ appear as a series of vertical strips. There is no contribution from ${\bf B}_x$ since ${\bf v_{\rm probe}}\times {\bf B}_x=0$. Therefore the density modulation recorded on the screen should appear as a series of vertical strips as in Fig. \ref{fig2}(c). In other words, the magnetic field topology inferred from the probe beam deflections is consistent with a helicoid as expected from the Weibel instability \cite{romanov_self-organization_2004}. We note that the data suggests that $B_{x,y}$ has radial dependence, which implies that there must exist $B_z$ component to satisfy $\nabla\cdot\bf B=0$. However, as shown in the Supplemental Material, $B_z$ is small near the axis of the plasma and therefore has negligible effects on the analysis below.

In Fig. \ref{fig2}(e), we show the density dependence of the magnetic field wavelength. Each data point is the average wavelength calculated using the integration of the time-resolved $k$ spectrum shown in the inset. The green curve shows the best fit to the data and gives the relation of $k=0.18\pm0.01~\omega_p/c$. For a collisionless plasma, the most unstable mode is $k_m=\sqrt{A/3}~\omega_p/c$ \cite{bret_multidimensional_2010}. This suggests that the plasma anisotropy plasma has dropped to a small value ($A<1$) when the signal become detectable in the experiment if one assumes $k\sim k_m$. As previously mentioned, such a rapid drop is attributed to precursor instabilities such as streaming and current filamentation instabilities as well as collisions \cite{zhang_ultrafast_2019}.

\begin{figure}
\includegraphics[width=0.45\textwidth]{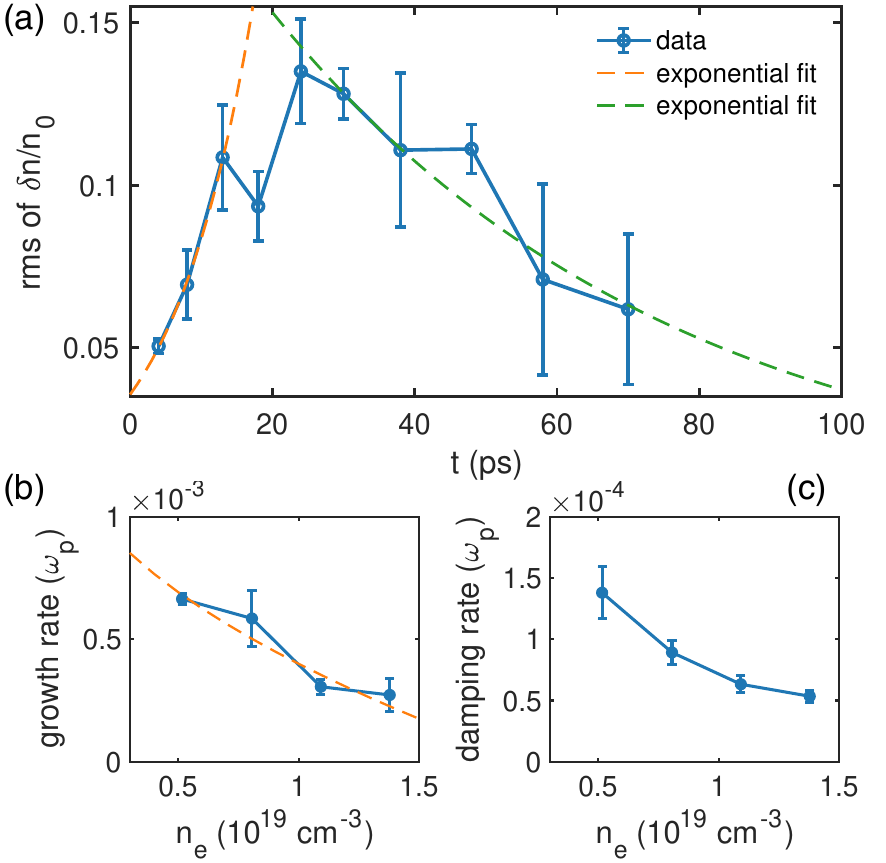} % this command will be ignored
\caption{(a) Temporal evolution of the measured density modulation. The error bar represents the standard deviation of multiple shots. (b) and (c), Density dependence of the deduced growth and damping rate of the signal. The error bar represents the standard deviation of the fitting coefficients in (a). The dashed line in (b) is the fit using the growth rate taking into account collisions ($\gamma_c=\gamma_0-\nu_e$, see text).}
\label{fig3}
\end{figure}

The temporal evolution of the deduced density modulation magnitude (Fig. \ref{fig2}(c)) is shown in Fig. \ref{fig3}(a). The data indicates a rapid growth, followed by peaking at around 20 ps, and then a slower decay. By fitting exponential curves to the data (the dashed lines) assuming a constant temperature anisotropy, we have extracted both the growth and damping rates, which are shown in Fig. \ref{fig3}(b) and (c), respectively. Without taking into account collisions, one would expect the normalized growth rate in Fig. \ref{fig3}(b) to be a constant since it is solely determined by $A$ and $T_\parallel$, which are determined by the precursor instabilities. Although the growth rates of the precursor instabilities do depend on plasma density, by the time the Weibel magnetic field becomes detectable the anisotropy level is about the same as evidenced by the constant measured $k\sim\sqrt{A/3}~\omega_p/c$ for different densities. It is known that collisions tend to reduce the growth rate, and narrow the width of the unstable spectrum towards smaller $k$ \cite{wallace_collisional_1987,mahdavi_role_2016}. The growth rate taking into account collisions is $\gamma_c=\gamma_0-\nu_e$ using the Krook’s collision model \cite{mahdavi_role_2016}, where $\gamma_0$ is the collisionless growth rate, $\nu_e=(1+Z)\nu_0$ is the collision rate that includes both the electron-electron collisions $\nu_{ee}=\nu_0\approx2.91\times10^{-6}n_e{\rm ln\Lambda} T_e^{-3/2}$ and the electron-ion collisions $\nu_{ei}=Z\nu_0$. Here $T_e=(2T_\perp+T_\parallel)/3$ is the effective electron temperature, $\rm ln\Lambda$ the Coulomb logarithm and $Z$ the charge state of the ions. Because of the dependence on plasma density of the collision rate, $\nu_e/\omega_p\propto\sqrt{n_e}$, the collisional growth rate $\gamma_c$ decreases with density. The time for electrons to traverse the width of the plasma has negligible effects on the instability growth compared to collisions (Supplemental Material).

The dashed line in Fig. \ref{fig3}(b) shows the best fit using the expression of $\gamma_c$ using the collisionless growth rate $\gamma_0=(1.5\pm0.3)\times10^{-3}~\omega_p$ and the effective electron temperature $T_e=230\pm50~\electronvolt$. Here the uncertainty represents the $1\sigma$ confidence level of the fitting coefficients. Since the normalized collisionless growth rate $\gamma_0/\omega_p$ is determined by the plasma temperatures $T_{\rm hot}=3(A+1)T_e/(2A+3)$ and $T_{\rm cold}=3T_e/(2A+3)$, we can further calculate the plasma anisotropy as being $A\approx0.48\pm0.08$ using $\gamma_0$ and $T_e$, which then implies $T_{\rm hot}\approx260~\electronvolt$ and $T_{\rm cold}\approx180~\electronvolt$. This anisotropy is consistent with the previous estimation based on the measured small wavenumber. The most unstable mode for these parameters is $k_m\approx0.38\pm0.04~\omega_p/c$ for a collisionless plasma, which is within a factor of two with the measured $k=0.18\pm0.01~\omega_p/c$, and the agreement is even better if one considers that collisions will reduce $k_m$. The measured effective temperature is lower than that obtained from 3D PIC simulations ($T_{e, \rm simu}\approx350~\electronvolt$, see details in \cite{zhang_probing_2020}) using the Osiris code \cite{Fonseca_OSIRIS_2002}. It can be attributed to the fact that in the experiment the low-temperature singly-ionized region of helium was larger due to the low-intensity wings of the laser spot which lowers the effective plasma temperature.

\begin{figure}
\includegraphics[width=0.45\textwidth]{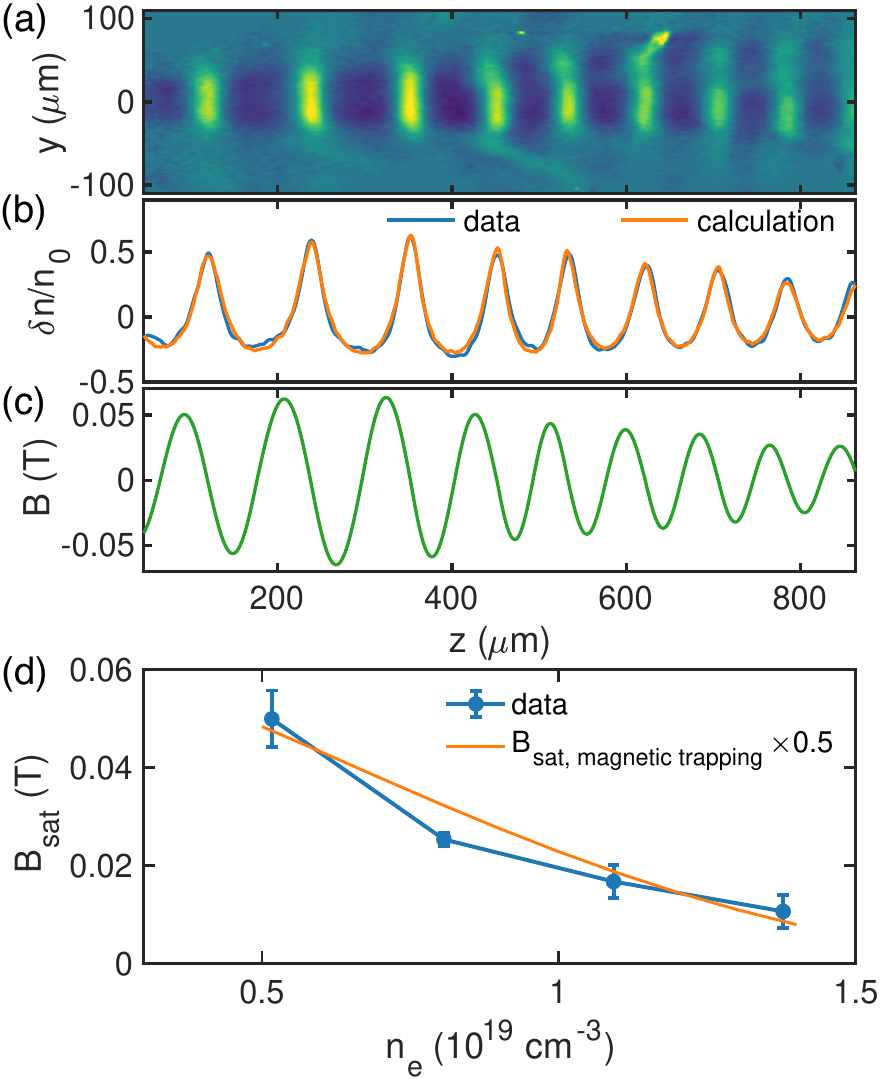} % this command will be ignored
\caption{(a) Measured density modulation when the instability reaches saturation (the same shot in Fig. \ref{fig2}(c) at t=24 ps). (b) The blue line shows the on-axis lineout of the density modulation whereas the orange line shows the calculated density modulation using the magnetic field shown in (c). (d) Saturated magnetic field amplitude as a function of plasma density. The orange line shows the estimated $B_{\rm sat}$ using the magnetic trapping mechanism (see text).}
\label{fig4}
\end{figure}

The isotropization of the plasma (aided by collisions) will terminate the growth of Weibel instability and eventually damp the magnetic fields. Figure \ref{fig3}(c) shows that the magnetic field damps more rapidly in the low-density plasma, indicating physical effects other than collisions play a role. The lifetime of a quasi-static but periodic magnetic field embedded in plasma (the so-called magneto-static mode \cite{lampe_interaction_1978,mori_generation_1991}) is of fundamental interest in plasma physics but it has not been possible to excite this mode in a plasma until now to our knowledge. Our measurements show that the saturated Weibel magnetic field is largely periodic although it can be spatially chirped (see Fig. \ref{fig4}(a)). This saturated state lasts for tens of ps (damping rate is $\sim10^{-4}~\omega_p$). Such a small damping rate implies that a magnetostatic mode generated by other methods \cite{mori_generation_1991,fiuza_high-brilliance_2010} may similarly last for a sufficiently long time, making these periodic magnetic fields useful as ultra-compact undulators.

In Fig. \ref{fig4}(a) we have replotted the measured $\delta n/n_0$ taken at $t=24$ ps in Fig. \ref{fig2}(c) when the instability had saturated. The blue line in Fig. \ref{fig4}(b) is the on-axis lineout of $\delta n/n_0$ and the orange line is the calculated density modulation using the magnetic field shown in Fig. \ref{fig4}(c). A parallel probe beam with the experimental parameters was propagated through a static magnetic field as in Fig. \ref{fig4}(c) and then tracked in vacuum for 23 cm to generate the calculated density modulation in Fig. \ref{fig4}(b). The magnetic field in the probe direction was simplified as being uniform with an FWHM width of $\sim70~\micro\meter$ estimated using the transverse size of the measured magnetic field in the orthogonal plane. Figure \ref{fig4}(c) shows that the amplitude of the saturated magnetic field is about $0.05~\tesla$. 

A well-known saturation mechanism of Weibel instability is magnetic trapping, which assumes that the instability saturates when the electron bouncing frequency in the magnetic field is on the same order of the growth rate. Following this assumption, the saturated magnetic field is expressed as $eB_{\rm sat}/m_ec\omega_p\sim \left<c/v_{\rm hot}\right>(\omega_p/kc)[\gamma(k)/\omega_p]^2$, where $\left<c/v_{\rm hot}\right>$ denotes the average over the particle distribution \cite{bret_multidimensional_2010}. Substituting the measured $k=0.18~\omega_p/c$ (Fig. \ref{fig2}(d)), growth rate (dashed line in Fig. \ref{fig3}(b)) and $v_{\rm hot}\approx0.02~c$ ($T_{hot}\approx260~\electronvolt$), we can calculate the saturated magnetic field amplitude as a function of plasma density, which is shown by the orange curve in Fig. \ref{fig4}(d). We note that the calculated $B_{\rm sat}$  has been multiplied by a factor of 0.5 to match the data. This factor implies that the Weibel instability saturates when the electron gyrofrequency reaches $70\%$ of the growth rate. This excellent agreement confirms the consistency of the data, namely, the relatively small magnetic field is due to the small growth rate at the small $k$, and $B_{\rm sat}$ varies inversely with plasma density due to the density-dependent growth rate caused by collisions. We note that alternative estimates of $B_{\rm sat}$ are also derived in literature (\cite{bret_multidimensional_2010} and references therein), but these estimates do not agree with the data (Supplemental Material).

In summary, we have made time-resolved measurement on ps time scale of the growth, saturation and damping of the electron thermal Weibel instability in underdense OFI plasmas using well-characterized relativistic electron bunches as a probe. The data is consistent with the magnetic fields self-organizing into a quasi-static helicoid structure thus confirming a long-standing prediction of the kinetic theory of the Weibel instability. The measured growth rates show density dependence that agree with the kinetic theory that takes into account collisions using the Krook model. The saturation mechanism is consistent with magnetic trapping. After saturation, the Weibel magnetic fields damp exponentially at a rate of $\sim10^{-4}~\omega_p$ and last for tens of ps with small change in its wavelength. In addition, the new probing technique we have demonstrated is suitable for exploring a broad range of plasma phenomena such as magnetic reconnection, annihilation and island formation occurring in magnetized plasmas and for studying astrophysical phenomena in the laboratory.

\begin{acknowledgments}
This work was supported by the Office of Naval Research (ONR) MURI (N00014-17-1-2075), AFOSR grant FA9550-16-1-0139, U.S. Department of Energy grant DE-SC0010064 and NSF grant 1734315. The work was also supported by the National Natural Science Foundation of China (NSFC) under Grants No. 11535006, No. 11991071 and No. 11775125. The authors thank Dr. Zheng Zhou and Hanxun Xu for helping with the experiment.
\end{acknowledgments}

\end{document}